\documentclass[journal = jctc,manuscript=article, layout=twocolumn]{achemso}

\usepackage{upgreek}
\usepackage{amsmath}
\usepackage{amsfonts}
\usepackage{amstext}
\usepackage{mathtools}
\usepackage{amssymb}
\usepackage{siunitx}
\usepackage[usenames]{color}
\usepackage[normalem]{ulem}
\usepackage{float}

\usepackage[utf8]{inputenc} 
\usepackage[T1]{fontenc}    
\usepackage{hyperref}       
\usepackage{url}            
\usepackage{booktabs}       
\usepackage{amsfonts}       
\usepackage{nicefrac}       
\usepackage{microtype}      
\usepackage{lipsum}
\usepackage{ulem}

\usepackage{amsmath}
\usepackage{booktabs}
\usepackage{subcaption}
\usepackage{graphicx}
\usepackage{xr}

\DeclareGraphicsExtensions{.pdf,.png}

\newcommand*{\diff}[1]{\mathop{\mathrm d #1}}

\newcommand*{\ka}{K$^+$ }
\newcommand*{\na}{Na$^+$ }
\newcommand*{\mean}[1]{\left< #1 \right>}
\newcommand*{\meanneq}[1]{\left< #1 \right>_\text{NEQ} }
\newcommand*{\NEQ}{_\text{NEQ}}
\newcommand*{\up}{\textsuperscript}

\newcommand*{\kT}{k_{\rm B}T}

\newcommand{\SIpmfDH}{S1}
\newcommand{\SItwoions}{S2}
\newcommand{\SItwoionsDH}{S3}
\newcommand{\SIpmfDHexit}{S4}
\newcommand{\SIdihedrals}{S5}
\newcommand{\SIdPCA}{S6}
\newcommand{\SIpaths}{S7}
\newcommand{\SIwork}{S8}
\newcommand{\SIpmfHtoH}{S9}
\newcommand{\SIpmfHtoHexit}{S10}
\newcommand{\SIpmfHtoHtwoions}{S11}
\newcommand{\SIchargedist}{S12}
\newcommand{\SIsmoothG}{S13}
\newcommand{\SILEtime}{S14}
\newcommand{\SILEmass}{S15}

\usepackage[fontsize=10pt]{scrextend}
\setkeys{acs}{articletitle = true}

\title{Predicting ion channel conductance via dissipation-corrected targeted molecular dynamics and Langevin equation simulations}
\author{Miriam J\"ager}
\affiliation{Biomolecular Dynamics, Institute of Physics, 
 University of Freiburg, 79104 Freiburg, Germany}
 \author{Thorsten Koslowski}
 \affiliation{Institute of Physical Chemistry, 
 University of Freiburg, 79104 Freiburg, Germany}
\author{Steffen Wolf}
\affiliation{Biomolecular Dynamics, Institute of Physics, 
 University of Freiburg, 79104 Freiburg, Germany}
\email{steffen.wolf@physik.uni-freiburg.de}

\begin{document}
\begin{abstract}
Ion channels are important proteins for physiological information transfer and functional control.
To predict the microscopic origins of their voltage-conductance characteristics,
we here applied dissipation-corrected targeted Molecular Dynamics in combination with Langevin equation simulations to potassium diffusion through the Gramicidin A channel as a test system. Performing a non-equilibrium principal component analysis on backbone dihedral angles, we find coupled protein-ion dynamics to occur during ion transfer. 
The dissipation-corrected free energy profiles correspond well to predictions from other biased simulation methods.
The incorporation of an external electric field in Langevin simulations enables the prediction of macroscopic observables in the form of I-V characteristics.

\end{abstract}
\maketitle


\section{Introduction}
Ion channels are membrane-spanning proteins that exist in every cell of every living organism\cite{Isacoff13}.
The function of these channels is to enable and control ion flux in and out of cells. The resulting charge currents lead to compartmentation and control of electrostatic gradients, which is one of the major mechanisms of information transfer within living beings\cite{Hille01, Catterall12}. Consequently, channel dysfunctions result in maladies such as central nervous disorders of excitability, e.g., epilepsy and cardiac arrhythmia\cite{shieh00}.

To gain insight into the molecular mechanisms of ion transfer through such channels, unbiased all-atom molecular dynamics (MD)\cite{Berendsen87} simulations can be used. This approach becomes impractical in the presence of high free energy barriers, leading to rare transitions that require long simulation times for statistically converged results. Furthermore, ion transport across channels is usually governed by external driving such as electrostatic and osmotic potentials.
To overcome such issues, various MD methods employing bias potentials 
have been applied to ion channels \cite{Roux93,Berneche01,Roux04a,Kato05,Bastug06,Khalili-Araghi09,Peng16,Na18,Flood19,Ngo21}, which usually require extensive equilibration. A way to circumvent equilibration is to carry out non-equilibrium simulations\cite{Liu06,deFabritiis08,Giorgino11}. 
Alternatively, simulations can be sped up by using coarse-graining approaches and electrostatic-based models\cite{Berneche01,Edwards02,Liu14,Wilson14}.

To implement external driving, it is possible to apply an electric potential along the simulation box\cite{Roux04,Roux08,Gumbart12,Koepfer14,Kopec19,Gu20}.

\begin{figure}[htb!]
	\centering
	\includegraphics[width=0.65\linewidth]{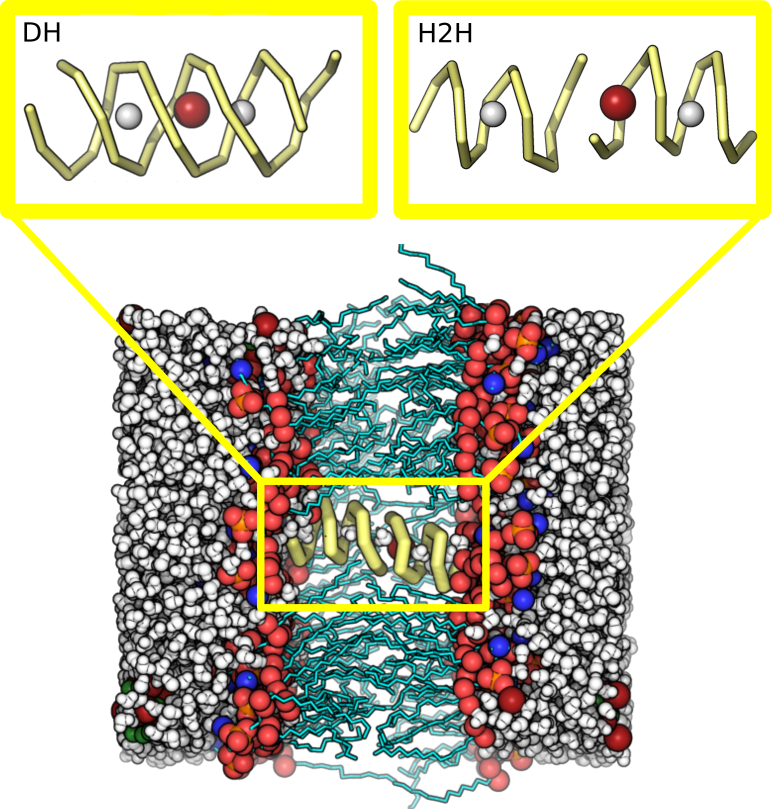}
	\caption{Simulation box of Gramicidin A (yellow) embedded in a DMPC lipid bilayer (cyan sticks and red / blue spheres) solvated by water (white) after equilibration. DH conformation and H2H conformation with \ka (red sphere) and pull group "anchors" (grey spheres).}
	\label{fig:gAbox}
\end{figure}

We recently developed an approach that combines non-equilibrium simulations and coarse-graining of system dynamics called dissipation-corrected targeted MD (dcTMD) \cite{Wolf18}, which allows to calculate free energies as well as friction profiles along a reaction coordinate of interest directly from a series of constant velocity targeted MD trajectories. Such free energies and friction factors can be used for the integration of a Langevin equation (LE), which allows to sample processes such as ion transduction well beyond the capabilities of atomistic MD simulations.\cite{Wolf20}.

We here use dcTMD in combination with LE simulations to calculate ion channel conductances. To mimic an electrophysiological experiment\cite{Busath98,Roux04}, we model the membrane potential by adding a linear potential to the free energy profile. From the resulting ion transition times we then calculate the corresponding ion currents. 
The major advantage of our approach is that we are not limited by the approximations made in rate theory,\cite{Kramers40,Roux91} but can calculate rates for systems with rugged free energy profiles and several transition states of similar height. Furthermore, we can predict I-V characteristics that do not obey Ohm's law \cite{Roux04}.
As test system, we use the Gramicidin A channel (gA) from \textit{Bacillus brevis} \cite{Andersen84,Wallace90}.
Gramicidines consist of a dimer of a 15 amino acid helical peptide and form a family of antibiotics, that damage and kill bacteria by increasing the cation permeability of their plasma membranes through bilayer-spanning pores. The cylindrical pore 
is only permeable for monovalent cations such as \na or \ka , but not for anions like Cl$^-$.
As shown in Fig.~\ref{fig:gAbox}, two main conformations of gA exist, which are the double helix (DH) and head-to-head (H2H) conformation. H2H is believed to be the physiologically relevant conformation\cite{CrossGramicidinControversy99, Andersen99} and is consistent with data from solid-state NMR spectroscopy\cite{ketchem96}, while DH appears in protein crystallization\cite{burkhart98} and organic solvents\cite{Burkhart99} (see the Supplementary Information for further details).
gA is both an experimentally \cite{Wallace90, Busath98, Edwards02,Saparov04,Andersen05, Kelkar07} and computationally \cite{Roux91b, Roux93, Elber95, deGroot02, Allen04, Kato05, Mamonov06, Bastug06, Liu06, deFabritiis08, Giorgino11, Peng16, Na18, Ngo21} well-investigated system, making it an ideal benchmark system for our approach.
We used both representative conformations as model systems for the simulations in this work:
the DH conformation served as theory-internal benchmark system, and the H2H conformation for comparison with other methods as well as with experiments.

\section{Theory}

\paragraph{Dissipation-corrected targeted Molecular Dynamics (dcTMD)}

We briefly recapitulate the theoretical basis of dcTMD: our approach is based on targeted MD developed by Schlitter et al. \cite{Schlitter94}. Here, a constraint force $f_c$ is applied to a subset of atoms to move it towards a target conformation along a predetermined one-dimensional path in conformational space with constant velocity $v_c$ 
along a pulling coordinate $s(t)=s_0+v_{c}t$.

From the resulting TMD trajectories, dcTMD \cite{Wolf18} estimates the free energy $\Delta G(s)$ as well as a non-equilibrium friction coefficient $\Gamma\NEQ$. dcTMD employs a second order cumulant expansion of Jarzynski's equality \cite{Jarzynski97, Jarzynski04}
\begin{align}
	\Delta G(s) = \meanneq{W(s)} - \frac{\beta}{2} \meanneq{\delta W(s)^2}
\label{eq:Jarzy}
\end{align}
where $\beta = 1/ \kT$, $W(s) = \int_{s_0}^{s} \text{d}s' f_c(s')$ denotes the work performed on the system by external pulling and $\meanneq{\cdot}$ an ensemble average over the independently realised pulling trajectories.
We further assume that the constraint force $f_c$ can be simply included as an additive term in a memory free Langevin equation \cite{Zwanzig01}
\begin{equation}
	m \ddot{s}(t) = - \frac{\text{d}G(s)}{\text{d}s} - \Gamma(s)\dot{s} + \sqrt{2 \beta^{-1} \Gamma(s)} \; \xi(t) + f_c(t).
	\label{eq:Langevin-noneq}
\end{equation}
with a mean force $-dG(s)/ds$, a dissipative drag force $ - \Gamma(s)\dot{s}$ and a Gaussian process $\xi(t)$ with zero mean and unity variance. As the constraint force $f_c$ enforces a constant velocity, $m\ddot{s} = 0$. An ensemble average of Eq.~(\ref{eq:Langevin-noneq}) over many TMD trajectories 
and integrating from $s_0$ to $s$ results in
\begin{align}
	\Delta G(s) &=  \meanneq{W} - v_c\int_{s_0}^{s} \text{d}s' \Gamma(s')
	\label{eq:dGGammaWdiss}
\end{align} 
where the 
second right-hand side term describes the dissipated work of the process in terms of the friction $\Gamma$. Combining Eqs. (\ref{eq:Jarzy}) and (\ref{eq:dGGammaWdiss}) finally yields non-equilibrium friction factors
\begin{equation}
	\Gamma_{\rm NEQ}(s(t)) = \beta \int_{t_0}^{t(s)} \diff{t'} \mean{\delta f_c(t) \delta f_c(t-t')}.
	\label{eq:GammaNEQ}
\end{equation}
Friction factors can be converted into diffusion coefficients $D = \kT / \Gamma$.

\paragraph{Path separation}
Eq.~(\ref{eq:Jarzy}) requires the assumption that the work along the pulling coordinate is normally distributed. 
However, the bias may introduce motion along additional hidden coordinates, leading to deviations from a normal work distribution\cite{Wolf20}. In the following, a "pathway" denotes a route through a relevant reaction coordinate space shared by a subset of trajectories. We showed earlier\cite{Wolf20} that clustering trajectories according to pathways and separately subjecting such clusters to dissipation correction reveals the free energies and friction profiles along those pathways. We assume that the most likely path taken is the one most energetically favoured, i.e. the one with the lowest free energy barrier. 

\paragraph{LE with external electrical field}
Using $\Delta G(s)$ and $\Gamma\NEQ$ estimated via Eqs.~(\ref{eq:dGGammaWdiss}) and (\ref{eq:GammaNEQ}) as input for the integration of the LE (\ref{eq:Langevin-noneq}), one can predict coarse-grained dynamics along $s$ as long as $\Gamma\NEQ\approx\Gamma_\text{EQ}$ \cite{Wolf20}. So far, this dcTMD-LE ansatz was only applied to systems without external driving except for the constraint force $f_c$. Here, we investigate its applicability to systems under external driving by an electric field. 
In a LE framework, this field can be represented by adding a linear electrostatic potential to the free energy. This approximation is valid if the electric field is homogeneous and stationary, only causes a linear perturbation of the free energy and has no influence on the system-bath time scale separation\cite{Zwanzig01}, i.e., the electric field does not alter the structure nor the dynamics of the channel. If these requirements are fulfilled, the resulting biased potential is
\begin{equation}	
	\Delta {\cal G}(s) = \Delta G(s) + \Phi (s)\cdot q.
	\label{eq:addU}
\end{equation}
with the electric potential $\Phi$ and ion charge $q$.
In difference to ion channels with charge sensitive domains \cite{Bezanilla08,StrutzSeebohm11,Delemotte15}, gA does not appear to perform conformational changes upon application of an electric field, and Eq.~(\ref{eq:addU}) should be well applicable for our investigation.

\section{Methods}

\paragraph{Structure preparation and equilibration}

gA models of a H2H dimer based on PDB ID 1MAG \cite{ketchem96} and an antiparallel DH structure based on PDB ID 1AV2 \cite{burkhart98} are based on simulation systems from Ref.~\cite{Na18}. 
The H2H as well as DH conformation had \ka placed at the preferred ion locations close to the center of the channel as determined in Ref.~\cite{Na18} and shown in Fig.~\ref{fig:gAbox}. 
The proteins were embedded into a bilayer patch of 115 DMPC lipids surrounded by ca. 3.600 TIP3P water \cite{Jorgensen83} with a 1~M concentration of KCl using the INFLATEGRO script \cite{Schmidt12}, resulting in a rectangular simulation box with dimensions of 6.05 x 6.05 x 6.64~nm. 

MD simulations were carried out in Gromacs v2016 and v2018 \cite{Abraham15} using a combination\cite{Cordomi12} of the Amber99SB force field\cite{Wang00,Hornak06} for the protein and the Berger force field\cite{Berger97,Tieleman99} for lipid parameters. 
Missing atomic parameters for N- and C-terminal modifications of gA were generated with antechamber\cite{Wang06} and acpype\cite{daSilva12} using GAFF atomic parameters \cite{Wang04} and AM1/BCC charges \cite{Jakalian00} used on a protocol applied by us before \cite{Wolf19a}.

For MD simulations, we used a 1~fs integration time step employing the leap-frog integrator \cite{Leach96}. Bonds between heavy atoms and hydrogen atoms were constrained by the LINCS algorithm \cite{Hess97}. Electrostatics were described by the
particle mesh Ewald (PME) method \cite{Darden93}. Cutoffs were set to 1~nm for van der Waals interactions and a minimum of 1~nm for PME real space. Temperature control was achieved by the
Bussi velocity rescaling thermostat\cite{Bussi07} (coupling time constant
of 0.2~ps). Pressure control was achieved via the Berendsen barostat \cite{Berendsen84} for preparation simulations employing positional restraints, and 
the Parrinello-Rahman barostat\cite{Parrinello81} for free MD simulations. In all cases, we used a semiisotropic pressure coupling with a coupling
time constant of 0.5~ps and a compressibility of 4.5x$10^{-5}$~bar\up{-1}. 

For structure equilibration, we employed a simulation protocol used by us previously\cite{Schneider11}: After an initial steepest descent minimization with positional restraints of 1000~kJ/mol applied to protein atoms and the bound \ka ion, 10~ns of MD simulations in the NPT ensemble retaining the restraints were carried out to pre-equilibrate the membrane-solvent environment. Afterwards, positional restraints were removed, and the full system was subjected to a second round of steepest descent minimization. After heating the system to 300~K in a short 10~ps simulation using positional restraints on protein atoms and the bound ion, restraints were removed again, and a final 10~ns free MD simulation for equilibration was carried out.

\paragraph{dcTMD simulations and pathway separation}

Targeted MD simulations \cite{Schlitter94} were carried out using the PULL code implemented in Gromacs \cite{Abraham15}. As pulling coordinate \textit{s} we used the distance between \ka and the center of mass (COM) of eight $C_\alpha$-atoms at the entrance of the channel (visualized in Fig. \ref{fig:gAbox}) that served as ''anchor group'' so that the pulling vector is parallel to the channel axis $z$.
During the simulation, \ka is pulled away or towards the anchor. 
As the PULL code only allows a maximum distance of half the shortest simulation box edge, the ion was pulled from to the exit of the channel in both directions to determine the free energy and friction profile along the whole channel. For better display of simulation data, the pulling coordinate $s$ was
mapped onto the channel $z$ axis using a Galileo transformation with the COM at $z=0$~nm. In the following, the two pulling directions are denoted as ''forward'' and ''backward'', indicating positive and negative values of $z$, respectively.

To generate an initial Boltzmann distribution following Eq.~(\ref{eq:Jarzy}), 100--1000 starting configurations with independent velocity distributions corresponding to a temperature of 300~K were generated from the equilibrated system after 10~ns free MD simulation. Each simulation system produced was then equilibrated for 10~ps with position restraints on protein atoms and the bound ion, followed by 100~ps of free MD simulation with a constant distance constraint applied to the ion--anchor group distance. Finally, pulling simulations with a constraint velocity $v_c = \SI{1}{m/s}$ were carried out in all systems for a simulation time of 1.2~ns, resulting in a cumulative simulated time of $\sim$4~$\upmu$s. This $v_c$ is close to a water permeation ''velocity'' of \SI{1.3}{m/s} in Gramicidin \cite{Pohl00} (see the Supplementary Information for details), and proved to be an optimal pulling velocity in dcTMD calculations on protein-ligand unbinding \cite{Wolf20}. 

The search for hidden reaction coordinates and pathways was performed with dPCA+ using the fastpca program \cite{Sittel17}. Trajectories were sorted according to pathways, and the resulting trajectory sets separately subjected to dissipation correction from Eqs.~(\ref{eq:dGGammaWdiss}) and (\ref{eq:GammaNEQ}) to obtain $\Delta G(s)$ and $\Gamma_{\rm NEQ}$. Friction profiles were smoothed with a Gaussian filter implemented in scipy\cite{Virtanen20} with a width $\sigma = 0.1$~nm (see Fig.~\SIpmfDH\ for the choice of this parameter).

\paragraph{Additional MD simulations}

To determine the bulk friction coefficient of potassium ions in our simulation setup, we performed dcTMD simulations on an enforced dissociation of a KCl ion pair in water. We used the same simulation protocol as in our study on the dissociation of NaCl\cite{Wolf18}, carrying out 1000 independent simulations of 1~ns each with a pulling velocity of $v_c = 1$~m/s. The bulk friction coefficient was determined as the average $\Gamma$ in the second half of the simulation.

For simulations with two potassium ions bound to gA, we substituted a water molecule at the positions of free energy minima revealed by dcTMD with a second potassium ion in both DH and H2H conformation with the first ion either at the channel entrance or in the center of the channel. After minimzation and equilibration using the protocol above, we carried out unbiased MD simulations of 10--20~ns length. 

For simulations with an electrical field applied to the DH conformation, we used the Gromacs option to add a homogeneous field to the simulation box. 
To achieve ion transitions within a reasonable simulation time ($\sim$300~ns), we applied a field of 0.08~V/nm corresponding to a voltage of 0.55~V over the length of the simulation box ($\sim$6.8~nm).
After equilibration using the protocol above, we carried out 20 independent MD simulations of each 100--300~ns individual length, with an accumulated simulation time of 3.6~$\upmu$s. While five simulations resulted into ion unbinding into the reverse direction, 12 simulations ended with a completed transfer. The remaining three simulations resulted in incomplete ion transfer and were included as a transfer duration of 300~ns into the calculation of the mean ion passage time.

\paragraph{Langevin equation simulations}

The biased potential ${\cal G}$ calculated via Eq.~(\ref{eq:addU}) as well as the friction profile $\Gamma$ derived from dcTMD were used as input for a Markovian LE equation \cite{Wolf20} (see the Supplementary Information for details).
The LE was numerically integrated using an integrator developed by Bussi and Parrinello \cite{Bussi07} using an integration time step of \SI{1}{fs}, the \ka mass $m=\SI{39}{g/mol}$ and $T=$\,\SI{300}{K}.
At least 100 LE simulations using different electric potentials $\Phi = 0, 0.01, ..., 0.3\;\text{V}$ were performed with a maximal simulation time of \SI{5000}{ns}. 
To compare simulation data based on the LE model to experimental values, the mean passage time $\tau_{\rm MPT}(\Phi)$ of \ka through the gA channel in dependence on the electric potential $\Phi$ was then calculated:
LE simulations started at $t_0$ close to the channel entrance $z=\SI{-1.2}{nm}$ and ended at a time $t_{\rm end}$ when the ion crossed $z=\SI{1.2}{nm}$. 
Single channel currents $I(\Phi)$ were then calculated as 
\begin{equation}
 	I(\Phi) = \frac{q}{\tau_{\rm MPT}(\Phi)} = \frac{q}{\mean{t_{\rm end}}(\Phi)}.
 \label{eq:currents}
\end{equation}
with the ion charge $q$. Uncertainties of $\tau_{\rm MPT}(\Phi)$ were derived using Jackknifing \cite{Efron81}.

\section{Results and Discussion}

\begin{figure}[htb!]
	\centering
	\includegraphics[width=0.95\linewidth]{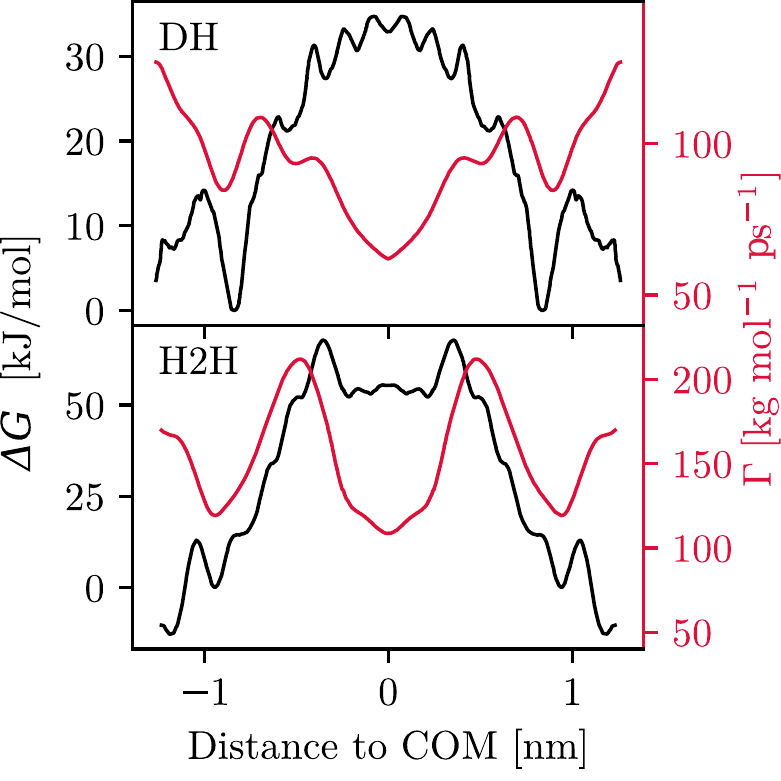}
	\caption{Final PMF (black) and smoothed friction profile $\Gamma$ (red) of \ka\ in gA in (A) DH and (B) H2H conformation.}
	\label{fig:h2hdhprofileplot}
\end{figure}

\paragraph{DH conformation}

From 100 TMD trajectories in each pulling direction, the free energy and friction profile $\Gamma(s)$ of \ka within the channel is recovered using dcTMD.
Following earlier works, the final profiles are symmetrized \cite{Allen06, Liu06} and shown in Figs.~\ref{fig:h2hdhprofileplot} and \SIpmfDH A. 
The global shape of the free energy profile exhibits a major central barrier with $\Delta G^{\neq} \approx \SI{35}{kJ/mol}$. Moreover, the profile exhibits two adjacent minima at $|z|=\SI{0.83}{nm}$ close to the channel entrance corresponding to binding sites, which fits observations from MD simulations in Ref.~\cite{Na18}. 


Further, we examined if the presence of two minima allows the binding of two ions at the same time. Short equilibrium MD simulations with two ions placed at these positions however led to the expulsion of one ion into the bulk solvent within 10~ns due to a perturbation of the single-file water chain (see Figs.~\SItwoions\ and \SItwoionsDH). The two free energy minima therefore only account for possible binding sites of a single ion present in the channel. Strictly speaking, the presented free energy profiles are only valid for the transfer of a single ion through gA in presence of a 1~M bulk KCl concentration. 
We note the possibility of ion transfer involving two ions, which we do not investigate here any further.

The friction profile displayed in Figs.~\ref{fig:h2hdhprofileplot} and \SIpmfDH B exhibits three pronounced minima: a global minimum in the middle of the channel, and two local minima at $\approx\SI{0.9}{nm}$ that approximately coincide with the binding site close to the channel entrance. 
The mean channel friction of $\mean{\Gamma_\text{DH}}=$ 93.5~kg/(mol~ns) corresponds to a diffusion coefficient $D_\text{DH} \approx 2.7 \cdot 10^{-3}$~\AA$^2$/ps for the DH conformation, which is about two orders of magnitude smaller than the one for potassium in bulk water ($\sim$0.6~\AA$^2$/ps).
Interestingly, the friction is on the order of values found for ligand unbinding from proteins \cite{Wolf20}, and similarly, maxima in friction are found at gradients in the free energy profile. This observation is in line with our earlier investigations \cite{Wolf18} that minima in free energies correspond to well-ordered states with only small fluctuations, and that approaching transition states leads to the disruption of such order and subsequently increased structural fluctuations. The global minimum around the center of the channel coincides with the maximum in free energy. 
After leaving the binding site the friction rises again, probably due to solvation of the \ka ion. Outside the channel, the estimate of the free energy becomes less reliable (see Figs.~\SIpmfDH\ and \SIpmfDHexit), but stays approximately constant within the estimated error range.

\begin{figure}[htb!]
	\centering
	\includegraphics[width=0.95\linewidth]{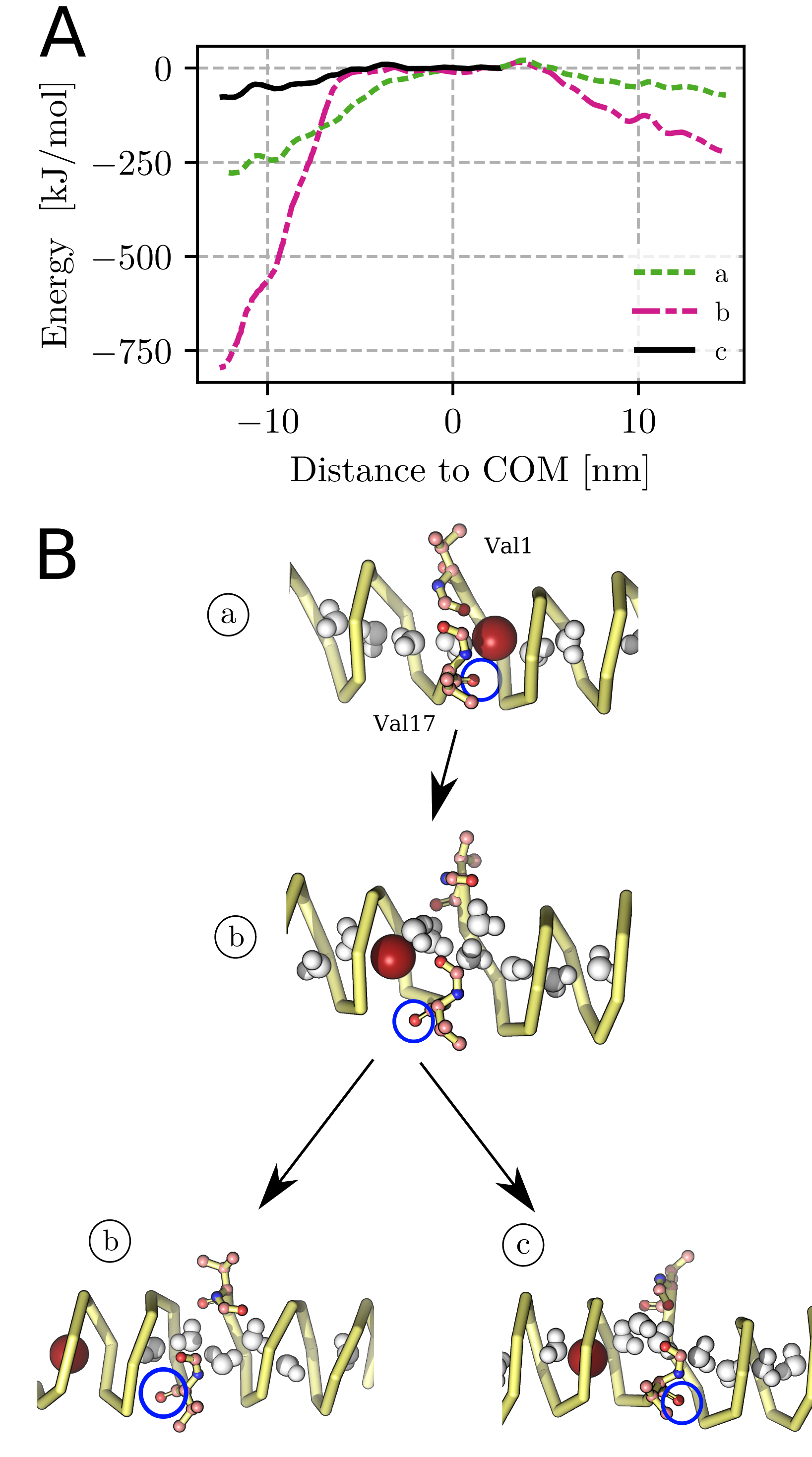}
	\caption{Path separation in the H2H conformation. (A) free energy profiles of different pathways and pulling directions described in the main text. (B) Snapshots of ion conduction in $-z$ direction. a) conformation at the beginning of the simulation. The Val1 carbonyl group (marked by blue ring) points towards \ka. Middle: Val17 follows \ka, causing the dihedral angle to flip. b) carbonyl group remains in its new position. This leads to Val1 pointing into the protein's center, 
		clogging the channel. c) carbonyl group flips back into its initial position, resulting a successful recovery of the channel geometry. }
	\label{fig:pathsepfigure2}
\end{figure}

\paragraph{H2H conformation}
For the analysis of this conformation, 1000 dcTMD simulations in each pulling direction were carried out. The starting position of \ka is at $z = 0.24$~nm, a position naturally taken by the \ka ion after equilibration. 
Applying our dissipation correction directly to the full set of trajectories resulted in free energy differences between simulation start and end on the order of several 100s of kJ/mol, which is about one order of magnitude higher than results from Umbrella Sampling calculations (29--\SI{50}{kJ/mol})~\cite{Na18}. Such artificial free energy profiles in dcTMD have been shown to arise from the presence of hidden coordinates \cite{Wolf20}, requiring a pathway separation via non-equilibrium PCA \cite{Post19}.

The small size of gA limits the number of possible candidates for such a hidden coordinate: as no large-scale conformational changes occur during ion translocation, we ruled out differences in protein-internal contacts\cite{Ernst17}. 
Instead, the H2H conformation is less rigid than the DH conformation in the middle of the channel, where the two dimers are connected by six hydrogen bonds around Val1 and Val17 \cite{Sun21}. Hence, changes of dihedral angles may occur during ion transfer, which in turn might affect degrees of freedom of the single file water chain inside Gramidicin \cite{deGroot02}. We therefore chose to perform a dPCA+ on the dihedral angles of gA (see Fig.~\SIdihedrals).
We find for the first principal component (PC1) that Val1 and Val17 indeed undergo conformational changes in the set of targeted MD simulations. Fig.~\SIdPCA A furthermore shows that only PC1 contains more than a single state along its biased energy \cite{Post19} $\Delta \mathcal{G}(\text{PC 1}) = - \beta^{-1} \ln \mathcal{P}(\text{PC 1})$, where $\mathcal{P}$ represents the probability to find the system at a given value of a PC within the set of biased trajectories.
Assessing changes of these dihedral angles in the simulations and further taking into account the stability of the single-file water chain, we find three distinct patterns of coupled dynamics of protein, ion and water chain that are visualized in Figs.~\ref{fig:pathsepfigure2}, \SIdPCA\ and \SIpaths\ (further details are given in the Suppleementary Information):

a) In most trajectories (730 trajectories in forward, 700 in backward direction), the valines remain in their initial conformation, and the water chain remains intact. This pattern is unproblematic when \ka is pushed out of the channel along $z \geq 0$~nm. However, when pushed in the opposite (backward) direction past Val1 and Val17, an artificial drop in free energy occurs again. 

b) In some trajectories (220 trajectories in forward, 70 in backward direction), the carbonyl group of Val17 follows \ka when the ion is pushed past the channel COM due to electrostatic interactions. This causes the corresponding dihedral angle to flip. As a results, Val17 "clogs" the channel and water molecules cannot follow \ka, leading to the local collapse of the protein structure and again a drop in free energy.  

c) In 200 trajectories in backward direction, the carbonyl group of Val17 follows \ka when the ion crosses the channel COM and then flips back into its initial position. Separating such trajectories and performing a separate dissipation correction leads to a free energy profile that agrees in shape and height with the one from a) in forward direction. 

The intermediate switching of Val17 in c) is consistent with observations from NMR experiments \cite{Jones10}, which propose a conformational change in the Val1/Val17 carbonyl group during ion transduction. We assume that a combination of the free energy profiles of a) for the simulations along $z>0$ and of c) for simulations along $z<0$ is the correct one. As can bee seen in Fig.~\SIwork, the resulting subsets of path-separated trajectories indeed result in the necessary recovery of a normal distribution of $\mean{W}$.
Possibly, the water chain collapsing in b) represents an artefact from the usage of a fixed charge force field, as protein-bound water chains are known to experience a significantly increased stabilization from polarization effects \cite{Wolf14a,Wolf14b}.

The resulting symmetrized PMF as well as the friction profile for the H2H conformation recovered from the path separated trajectories is shown in Figs.~\ref{fig:h2hdhprofileplot} and \SIpmfHtoH A. The qualitative shape of the free energy agrees with the one from the DH conformation: minima and maxima are approximately at the same positions and agree in depth and height, respectively. The profile exhibits two local minima inside the channel at $|z|\approx\SI{0.9}{nm}$, which match a binding site observed in experiment \cite{Olah91} as well as computational study \cite{Na18}. The free energy further decreases towards the exit, pointing towards a second binding site right outside the channel as observed by other studies\cite{Allen04, Allen06, Peng16}, and then roughly stays constant within the error of the estimate (see Fig.~\SIpmfHtoHexit). 
As in the case of the DH conformation, placing a second potassium ion into the second minimum leads to the exit of one of the ions (see Figs.~\SItwoions\ and \SIpmfHtoHtwoions) due to a perturbation of the water chain. Likewise, the two minima are only valid for a single ion passing through the channel.
Two free energy barriers of $\Delta G^{\neq} =\SI{68}{kJ/mol}$ can be found at $|z|\sim\SI{0.4}{nm}$, which is in the range of results from other targeted MD studies (58-79~kJ/mol) \cite{Giorgino11, deFabritiis08}. In other biased MD simulations using mainly Umbrella sampling methods, the obtained free energy barriers of gA in H2H conformation were found to be between $\approx 29-50\;\text{kJ/mol}$ \cite{Allen06, Na18}, which is at least \SI{18}{kJ/mol} lower than our result. In the center of the channel, a high-energy plateau can be found instead of a single maximum as transition state, which qualitatively matches the results from a recent study employing polarizable force fields \cite{Ngo21}.

As in the case of the DH conformation, the friction profile displayed in Fig.~\SIpmfHtoH B is minimal at the center of the channel, and peaks at gradients of the free energy profile. The friction rises from $|z|=\SI{0.2}{nm}$ until it peaks at $|z|\approx\SI{0.5}{nm}$ and decreases towards the channel entrance binding sites, and qualitatively agrees with the diffusion constant profiles presented in Ref.~\cite{Peng16}. 
The mean friction $\mean{\Gamma_\text{H2H}}\approx$~150~kg/(mol~ps) corresponds to a mean diffusion coefficient of $D_\text{H2H} \approx 1.6 \cdot 10^{-3}$~\AA$^2$/ps, which agrees with other investigations \cite{Peng16} within a factor of $\sim$10, and is well comparable to values obtained in similar nonequilibrium simulations employing steered MD\cite{deFabritiis08} ($\sim$1.8$ \cdot 10^{-3}$~\AA$^2$/ps).


\paragraph{LE with electrical field}

\begin{figure}[htb!]
	\centering
	\includegraphics[width=\linewidth]{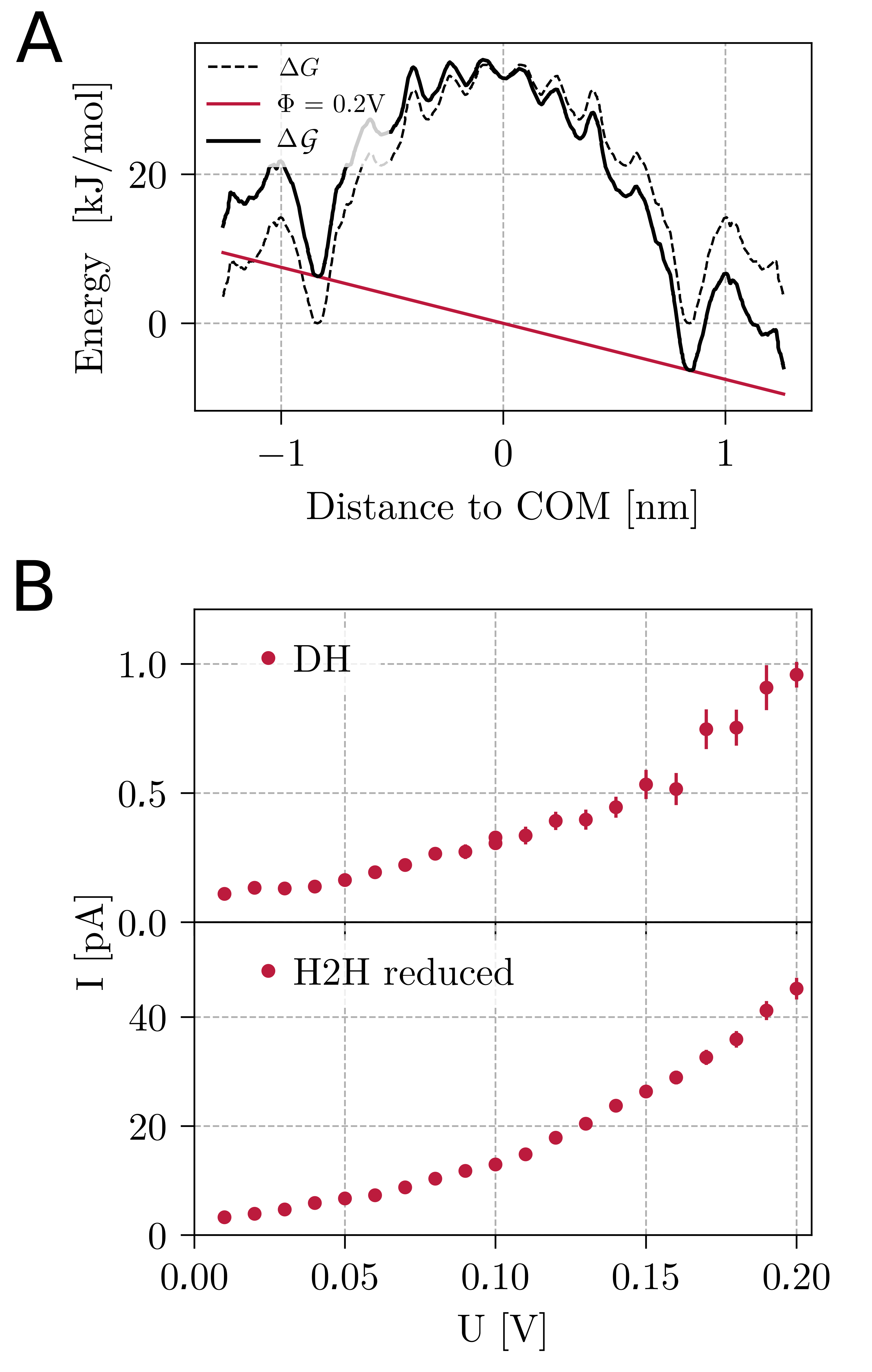}
	\caption{A: example $\Delta {\cal G}(z)$ (bold line) of DH at $\Phi=\SI{0.2}{V}$. The electric potential (red line) is added to $\Delta G$(dashed line) which was using dcTMD. B: resulting I-V curves from LE simulations. }
	\label{fig:i-v-curves}
\end{figure}

To ensure that the addition of a linear electric potential to the free energy in LE simulations is justified,  
we checked the charge distribution along the DH in simulations with and without electric field. As can be seen in Fig.~\SIchargedist A, the charge distributions differ only slightly, and mostly agree within 1$\sigma$.
Figure~\ref{fig:i-v-curves}A displays the change of the free energy curve by an electric potential following Eq.~(\ref{eq:addU}), and Fig.~\SILEtime\ displays typical distributions of mean passage times $\tau_{\rm MPT}$ resulting from LE simulations. 
Table~S1 compares our predicted $\tau_{\rm MPT}(\Phi)$ with values from experiment \cite{Busath98}, and Fig.~\ref{fig:i-v-curves}B displays selected current-voltage (I-V) characteristics calculated according to Eq.~(\ref{eq:currents}) for both the DH and the H2H conformation.

In fully atomistic simulation of the DH conformation with electric field, we observe $\tau_{\rm MPT, MD}= 138\, \pm \, 29$~ns. A comparison with the results from our LE simulations however is not straightforward: as Fig.~\SIchargedist B shows, the water-membrane interface where the linear potential drop occurs\cite{Roux08,Gumbart12} is not easy to define. Additionally, the channel with its length of $\sim$2.4~nm does not penetrate the membrane completely, but opens two inlet funnels. We therefore regard three likely scenarios: the potential drop starts and ends

\begin{enumerate}
\item[A:] at the beginning of the membrane density at $\pm$2.5~nm distance to the channel COM, resulting in a partial 0.27~V over the channel.
\item[B:] at the maxima in membrane density at $\pm$1.5~nm distance to the channel COM, resulting in a partial 0.44~V over the channel.
\item[C:] at the channel entrance and exit, resulting in the full 0.55~V over the channel.
\end{enumerate}

If we take scenario B as the most likely one and the two others as an upper and lower boundary, respectively, Tab.~S1 shows that we obtain an agreement of fully atomistic MD and LE simulation within a factor of $\sim$8$\pm_{6}^{8}$. These factors are well within the capabilities of our dcTMD-LE combination reported before\cite{Wolf20} and within the best error range that can be achieved by free energy-based methods\cite{Bruce18,NunesAlves20,Schindler20}. We think accordingly that our combined dcTMD-LE simulations approach gives good results for the reproduction of ion transition rates from MD simulations.


For the H2H conformation, \ka does not cross the channel within \SI{5000}{ns} in any of the LE simulations, which we attribute to an overestimation of free energy barriers. PMF calculations with the H2H system employing polarizable force fields \cite{Peng16,Ngo21} suggest that free energy barriers from simulations with fixed charge force fields are significantly too high. 
We therefore rescaled the H2H free energy profile to a maximal height of $\Delta G^{\neq}=\SI{20}{kJ/mol}$ as found in Ref.~\cite{Peng16}. 
The resulting I-V characteristics shown in Fig.~\ref{fig:i-v-curves}B agree in their exponentially increasing shape with measurements \cite{Busath98} for KCl concentrations of 1--2~M and only deviate from the absolute currents by a factor of~$~\sim$10, which is within the best range achievable by our method \cite{Wolf20}. 
We note that for smaller ion concentrations, experimental curves are known to exhibit an apparent linear or asymptotically increasing shape. 
As our LE simulations start when the ion just entered the channel at \SI{-1.2}{nm} and end at \SI{1.2}{nm} just prior to leaving the channel, the estimated $\tau$ and I-V curves should only be quantitatively comparable to electrophysiological experiments at high ion concentration, where the ion current is not limited by the ion diffusion rate from the bulk into the channel. 

Concerning the friction profile employed in LE simulations, we find that the transition times and calculated currents are not sensitive to the chosen smoothing factor (see Fig.~\SIsmoothG). However, using a potassium bulk friction coefficient leads to reduced currents in H2H and a complete vanishing of currents in DH. The latter observation highlights the necessity to use the dcTMD-derived channel friction coefficients for LE simulations. The drop in currents upon using a smaller friction coefficient stands in contrast to Kramers rate theory, where the transfer rate is $k \sim 1/\Gamma$, and smaller $\Gamma$ should result in a larger rate and thus increased currents. 

We note in passing that we carried out LE simulations with a comparatively small integrator time step of 1~fs, but the $\tau_{\rm MPT}$ appear to be insensitive to increasing the step size (see Fig.~\SILEtime). We attribute the time step insensitivity in both protein conformations to the existence of several smaller barriers with individual heights well beyond $\kT$ instead of a single transition state. Overcoming one such barrier due to an overlong time step thus does not directly result in a successful ion transfer. However, changing the system's mass as done in earlier investigations on protein-ligand pair dissociation \cite{Wolf20} resulted in significantly changed $\tau_{\rm MPT}$ (see Fig.~\SILEmass). This again points to the gA-ion system not following a Kramers reaction-rate expression in the high-friction regime\cite{Kramers40}. Interestingly, this finding is in contrast to the aforementioned work on protein-ligand complexes despite a comparable magnitude of $\Gamma$.

\section{Conclusion}

Our investigation of potassium conduction through Gramicidin A using dcTMD and LE simulations under driving by an external electric potential shows promise as a tool to explicitly compute single channel I-V curves from molecular simulations that can be directly compared to their experimental counterparts.
Our method requires full-atom simulations of $\sim$2.5~$\upmu$s total simulation time per channel conformation, which is comparable to other methods employing biased simulations\cite{Ngo21}. When taking into account polarization effects, the predicted absolute currents are within a factor of $\sim$10 of experimentally measured currents, which corresponds to an error in the transition barrier of $\sim$2~$\kT$ and is within the best range achievable by biased simulations\cite{NunesAlves20,Wolf20,Schindler20}. While most other methods underestimate the maximal conductance at 100~mV by at least a magnitude (see Tab.~1 in Ref.\cite{Peng16}), we overestimate it by a magnitude owing to the neglect of the initial ion transfer from the water bulk into the channel and the final transfer back into the bulk. A comparison of diffusion coefficients between different methods is problematic, as they can differ between methods by several orders of magnitude (see Ref.\cite{Mamonov06}). We find that our H2H mean diffusion coefficient matches the ones from similar nonequilibrium simulations \cite{deFabritiis08}.
Our approach allows us to monitor coupled ion-protein and ion-water dynamics during the transit as observed in experiment\cite{Jones10} and map them to the free energy and friction profiles. We note that studies using non-equilibrium steered MD\cite{Liu06, deFabritiis08, Giorgino11} have a similar capability to reveal such dynamics. Furthermore, including an electrical field in our Langevin simulation approach allows to directly calculate electrical currents in form of an I-V curve and therefore a direct comparison with experiment. As the Langevin simulations are extremely fast (1~$\upmu$s simulation take ca.~4 minutes on a single CPU), we can calculate hundreds of replicate runs at a range of voltages within a short time, and obtain well-converged current estimates. 


Concerning the comparison of I-V characteristics shape with experiment, we expect that any decrease in current at lower ion concentration than the 1~M investigated here will simply be a consequence of a first-order reaction law for ion binding to the channel entrance. At high ion concentrations on the other hand, transfer over the channel becomes the rate-limiting step. We cannot rule out a transfer mechanism involving two or more ions, which we did not investigate here any further.

The major challenge we encounter in H2H simulations is the significant underestimation of currents, which has been observed by others as well\cite{Allen06, Jensen13,Furini20}. The reason for this effect is an overestimation of free energy barriers along the channel due to a missing polarization term in the force field utilized here \cite{Allen06, Peng16,Ngo21}. 
A similar issue has recently been reported for nonequilibrium simulations of ligand unbinding from receptor molecules \cite{Capelli20}. Further application of the dcTMD-LE simulation combination to ion channels may therefore require the usage of polarizable force fields.

Finally, we need to address some differences between the applicability of dcTMD to gA and other physiologically relevant ion channels. First, 
using gA as test systems has the benefit of the protein conformation being independent from the applied electric potential. Other physiologically relevant ion channels such as potassium channels \cite{Berneche01,Koepfer14,Kopec19,Gu20} usually contain voltage-sensor-domains \cite{StrutzSeebohm11,Delemotte15} that cause an activation or inactivation of the channel via conformational changes depending on the applied voltage. However, if the respective open and closed states can be structurally identified, our approach may help in discriminating and explaining differences in conformation state conductivity.
Second, the symmetrization of free energy profiles is only possible in gramicidin due to its structural symmetry, and has been employed by others as well\cite{Allen06,Giorgino11,Peng16,Ngo21}. For the application to physiologically relevant ion channels, we will need to further evolve our method.
Last, such channels usually contain several potassium ions within the channel at the same time\cite{Berneche01,Koepfer14,Kopec19,Gu20}, which will necessitate a different bias coordinate as well as a careful evaluation of the number and position of ions and water molecules conducted.

\section{Associated content}
\subsection{Supporting Information}
One PDF with
(1) Supplementary Methods detailing on simulations with two ions in gA, pathway separation in H2H and Langevin simulations.
(2) Supplementary Discussion on gA conformation, gA water and ion conduction measurements, and work distribution analysis.
(3) One Supplementary Table.
(4) Fifteen Supplementary Figures.

This information is available free of charge via the Internet at \url{http://pubs.acs.org}.

\section{Acknowledgements}

The authors are grateful to Gerhard Stock, Matthias Post, Sehee Na, Lucie Delemotte and Wojciech Kopec for helpful discussion, and to Benjamin Lickert for discussions and help with Langevin equation simulation scripts. This work has been supported by the Deutsche Forschungsgemeinschaft
(DFG) via grant WO 2410/2-1 within the framework of the Research Unit FOR 5099 ''Reducing complexity of nonequilibrium'' (project No. 431945604). The authors acknowledge support by the bwUniCluster computing initiative, the High Performance and Cloud Computing Group at the Zentrum f\"ur Datenverarbeitung of the University of T\"ubingen, the state of Baden-W\"urttemberg through bwHPC and the DFG through grant No.~INST 37/935-1 FUGG.

\providecommand{\latin}[1]{#1}
\providecommand*\mcitethebibliography{\thebibliography}
\csname @ifundefined\endcsname{endmcitethebibliography}
  {\let\endmcitethebibliography\endthebibliography}{}


%

\end{document}